\RequirePackage{fix-cm}
\documentclass[smallextended]{svjour3}       
\smartqed  
\usepackage{graphicx}
\usepackage[utf8]{inputenc}
\usepackage{amsmath}
\usepackage{amssymb}

\begin{document}

\title{Uncertainty relation and inseparability criterion}

\author{Ashutosh K. Goswami$^1$        \and
        Prasanta K. Panigrahi$^1$ 
}


\institute{ \at
           \email{ashutoshgoswami841@gmail.com}
            \at
          \email{pprasanta@iiserkol.ac.in}\\ 
          $^1$ Indian Institute of Science Education And Research Kolkata, Mohanpur - 741 246, West Bengal, India}

\date{Received: date / Accepted: date}

\maketitle

\begin{abstract}

  We investigate the Peres-Horodecki positive partial transpose (PPT) criterion in the context of conserved quantities  and derive a condition of inseparability for a composite bipartite system depending only on the dimensions of its subsystems, which leads to a bi-linear entanglement witness for the two qubit system. A separability inequality using generalized  Schrodinger-Robertson uncertainty relation taking suitable operators, has been derived, which proves to be stronger than the bi-linear entanglement witness operator.  In the case of mixed density matrices, it identically distinguishes the separable and non separable Werner states.  
\end{abstract}

\section{Introduction}
Entanglement $[1]$ is one of the unique features of quantum mechanics, not present in classical physics.   Einstein, Podolsky and Rosen (EPR) described this quantum correlation as "spooky action at a distance", in their well known paper ${[2]}$.  Presently, entanglement is recognized as a resource for many quantum information processing tasks such as, teleportation ${[3]}$, super dense coding $[4]$, quantum information splitting $[5]$, quantum cryptography $[6]$ and measurement based quantum computation $[7]$ to mention a few.\\

Entanglement characterization for  multipartite system is far from being understood.  For $2\otimes2$ and $2\otimes3$ systems Peres-Horodecki $[8, 9]$ positive partial transpose (PPT) criterion provides a complete characterization of entanglement.  The PPT criterion is the sufficient condition for entanglement, stating that if a bipartite state $\rho$ is separable, it can be written as $\rho = \sum_i \rho^a \otimes \rho^b$ and its partial transpose $\rho^{pt} = \sum_i \rho^a \otimes \rho^{b^T}$ must be a valid density matrix (positive). The same has been extended to continuous variable Gaussian states by Simon $[10]$, Duan et al. $[11]$, where PPT manifests as the operator relation,\\ 
\begin{equation}
 (x_{a},p_{a},x_{b},p_{b})\rightarrow (x_{a},p_{a},x_{b},-p_{b}). 
\end{equation}
Here $x_{a}$, $x_{b}$ and $p_{a}$, $p_{b}$ are the position and momentum operators for the two particles a and b, respectively.  When implemented on uncertainty relation of EPR type variables, PPT leads to a separability condition.  More precisely, EPR variables, $u=\frac{x_{a}+x_{b}}{2}$ and $v=\frac{p_{a}+p_{b}}{2},$ satisfy the commutation relation,
$[u,v]$= $\imath$, $[\hbar=1]$, with corresponding uncertainty relation, \\

\begin{equation}
  \Delta(u)\Delta(v)\geq\frac{1}{2}.
\end{equation}
Under PPT, the following counter intuitive relation for the pair of commuting observables is obtained for the separable states,
\begin{equation}
\Delta(\frac{x_{a}+x_{b}}{2})\Delta(\frac{p_{a}-p_{b}}{2})\geq \frac{1}{2}.   
\end{equation}
Violation of the above inequality provides a sufficient condition for entanglement.
The implication of Peres-Horodecki  positive partial transpose (PPT) criterion for the non-Gaussian states has been studied by Agarwal and Biswas $[12]$, using Heisenberg uncertainty relation and generalised  Schrodinger-Robertson uncertainty relation (SRUR) $[19]$ by H.  Nha $[13].$  Observables possessing underlying SU(2) and SU(1,1) algebraic structures, important from the quantum optics point of view, have been investigated for bipartite Hilbert space  and non-Gaussian state has been cast as a finite dimensional system and the need of higher order momenta to characterize entanglement of these non-Gaussian states is shown $[12]$. The inseparability inequality for finite dimensional quantum systems has been studied in $[14-18]$.  In ref. $[15]$ observables used satisfy,  a constraint $(A^2)^{pt} = (A^{pt})^2$ $[15]$, which detect various cases of discrete and continuous variable entanglement.  Further, it is shown  that such constraints are not necessary for this purpose $[16]$.  Here, we investigate PPT criterion in the context of conserved quantities  and derive a condition of inseparability for a composite bipartite system depending only on the dimensions of its subsystems,  which leads to a bi-linear entanglement witness $[20]$ in the case of two qubit systems , obtained earlier.  Subsequently, a non-linear inseparability inequality using the condition of separability $[16]$ with the suitable operator commutation relation $[A,B]=\imath C$ has been derived, which proves to be stronger than a bi-linear entanglement witness operator and it combines two bi-linear entanglement witness operators in a single witness.  For the Werner mixed state case, the given inequality detects entanglement when $x>\frac{1}{3}$, the exact condition for  entanglement obtained through PPT $[9]$, thus improves the result of $[14, 15] (x>\frac{1}{\sqrt{3}})$ and $(x>\frac{1}{2})$, respectively and the Bell inequalities $[21]$ $(x> \frac{1}{\sqrt{2}})$.\\
For the sake of clarity, we will prove the following results that will be used throughout the paper;
\paragraph{\textbf{Lemma 1.}}
The relation $tr(M\rho^{T}) = tr (M^{T} \rho)$ holds for any density matrix $\rho$ and any observable M.\\
Proof: Writing the components of the matrix as $M_{ij} =<i\mid M\mid j>$ in the basis $\big\{ \mid i> \big\}$ and using the fact that transposition of a matrix swaps the indices i.e., $M^T_{ij} = M_{ji}$, we have $tr(M \rho^{T}) =  M_{ij} \rho_{ij} =  tr (M^{T} \rho)$, where repeated indices are summed up.
\paragraph{\textbf{Cor. 1.}}  Under transposition Pauli matrices transform as, $(\sigma_{x}, \sigma_{y},\sigma_{z}) \rightarrow  (\sigma_{x}, -\sigma_{y},\sigma_{z})$\\
Proof: Using Lemma $1$, $tr( M \rho^{T}) = tr (M^{T} \rho)$ and relations, $<M>_\rho = tr(M\rho)$ and  $\sigma_x^{T} = \sigma_x$, $\sigma_y^{T} = -\sigma_y$ and $\sigma_z^{T} = \sigma_z$, we have  $ <\sigma_x>_{\rho^T} = <\sigma_x>_{\rho}$, $<\sigma_y>_{\rho^T} = -<\sigma_y>_{\rho}$ and $<\sigma_z>_{\rho^T} = <\sigma_z>_\rho$, which proves the corollary.  It is noted that lowering and raising operators transform as, $$a=(\frac{\sigma_{x} + i \sigma _{y}}{2}) \leftrightarrow a^{\dagger}=(\frac{\sigma_{x} - i \sigma _{y}}{2}),$$ as in continuous variable case $[10].$
\paragraph{\textbf{Lemma 2.}} 
The relation $tr(M\rho^{pt}) = tr (M^{pt} \rho)$ holds for any density matrix $\rho$ and any observable M.\\
Proof:  Writing the components of the matrix as $ M_{i\alpha,j\beta} = <i,\alpha\mid M \mid j,\beta> $, where $\big\{\mid i>\big\}$ is the basis of the first subsystem and $\big\{\mid \alpha>\big\}$ is the basis of the second subsystem and using the fact that under partial transpose indices $\alpha$ and $\beta$ interchange i.e., $M_{i\alpha,j\beta}^{pt} = M_{i\beta,j\alpha}$, we have $tr(M\rho^{pt}) =  M_{i\alpha,j\beta} . \rho_{j\alpha,i\beta} = tr (M^{pt} \rho).$
\paragraph{\textbf{Cor. 2.}} Under partial transposition, Pauli matrices for two subsystem transform as,\\ 
 $$(\sigma_{x}, \sigma_{y},\sigma_{z})_{a} \rightarrow  (\sigma_{x}, \sigma_{y},\sigma_{z})_{a}, (\sigma_{x}, \sigma_{y},\sigma_{z})_{b} \rightarrow (\sigma_{x}, -\sigma_{y},\sigma_{z}) _{b}.$$\\
  Proof: Using Lemma $2$,
\[
      <\sigma_i \otimes \sigma_j>_{\rho^{pt}} = 
\begin{cases}
    -<\sigma_i \otimes \sigma_j>_\rho, & \text{if } j = y\\
    <\sigma_i \otimes \sigma_j>_\rho,              & \text{otherwise},
\end{cases}
\] which proves the corollary.

We consider bipartite spin system of j $\otimes$ j' dimension.  The eigenstates of the composite system are characterized by the invariants $\vec{S}^2$, $S_{z}$, $\vec{S_{a}}^2$,  $\vec{S_{b}}^2$.  Here, $\vec{S}=\vec{S_a + S_b}$ and $S_{z}=S_{a_z}+S_{b_z}$ are the total angular momentum and z-component of the total angular momentum operator, respectively.  The quadratic casimir invariant is, $\vec{S}^{2}=(\vec{S_{a}} + \vec{S_{b}})^{2} = \vec{S_{a}}^{2} + \vec{S_{b}}^{2} + 2 \vec{S_{a}}.\vec{ S_{b}}$, where $\vec{S_a^2} = j(j+1)$, $\vec{S_b^2}=j'(j'+1)$ and  $\vec{S_{a}}.\vec{ S_{b}}=S_{a_x} S_{b_x}+S_{a_y} S_{b_y}+S_{a_z} S_{b_z}.$ The eigenvalues of the  $\vec{S}^2$ are s (s+1) with allowed values of s (assuming j$\geq$ j', without loss of generality),
$$t, t-1, t-2. \dots t_i \dots t',$$ where t=j+j' and t'=j-j'. Corresponding eigenvalues of $P = \vec{S_{a}}.\vec{ S_{b}}$ are,
$$ jj', \dots,P_i,\dots -j'(j+1). $$
The eigenvalues of $\vec{S_{a}}.\vec{ S_{b}}$ are bounded above by jj' and below by -j'(j+1).  For any superposition state $\mid\psi> = \sum_i c_i \mid t_i>$, the expectation value of the operator $\vec{S_{a}}.\vec{ S_{b}}$ is given by, $$<\vec{S_{a}}.\vec{ S_{b}}> = \sum_i \mid c_i\mid^2 (\vec{S_{a}}.\vec{ S_{b}})_i;\quad \mid c_i\mid^2 \leq 1,\quad \sum_i \mid c_i\mid^2 = 1.$$ The set of real numbers is a convex set, hence $<\vec{S_{a}}.\vec{ S_{b}}> \in  [-j'(j+1),jj'].$ Similarly, for any density matrix $\rho = \sum_i p_i\mid \psi_i><\psi_i\mid$, where $p_i \leq 1$ and $\sum_i p_i = 1,$
 \begin{equation}
  <\vec{S_{a}}.\vec{ S_{b}}> \in [-j'(j+1),jj']  
\end{equation} 
Under partial transpose  operation, $$\vec{S_{a}}.\vec{ S_{b}} \rightarrow (\vec{S_{a}}.\vec{ S_{b}})_{pt}.$$ If the given state $\rho$ is separable, following condition must be satisfied, \\
    \begin{equation}
        <(\vec{S_{a}}.\vec{ S_{b}})_{pt}> \in [-j'(j+1),jj'].
    \end{equation}
For the two qubit systems $(j=\frac{1}{2})$, $\vec{S_{a}}.\vec{ S_{b}}=\frac{1}{4} ( \sigma_{x} \otimes \sigma_{x} +\sigma_{y} \otimes \sigma_{y}+\sigma_{z} \otimes \sigma_{z})$  and  $(\vec{S_{a}}.\vec{ S_{b}})_{pt}=\frac{1}{4} ( \sigma_{x} \otimes \sigma_{x} -\sigma_{y} \otimes \sigma_{y}+\sigma_{z} \otimes \sigma_{z})$, hence the 
 condition of separability in terms of Pauli matrices is obtained as, 
   
 $$ G=<\vec{\sigma_{a}}.\vec{\sigma_{b}}>_{\rho^{pt}}=<\sigma_{x_a} \sigma_{x_b} - \sigma_{y_a} \sigma_{y_b}+\sigma_{z_a} \sigma_{b_z}>_{\rho} \in [-3,1] $$  
A bi-linear entanglement witness operators can be obtained from above condition, $I - \sigma_{x} \otimes \sigma_{x} +\sigma_{y} \otimes \sigma_{y}-\sigma_{z} \otimes \sigma_{z} \geq  0,$ also recognized as teleportation witness earlier $[22].$

\subsection{Inequality from the generalized operator algebra $[A,B]= \imath C$}
For any observables A, B, having commutator $[A,B]=\imath C$, the SRUR leads to, 

\begin{equation}
 \Delta(A)^2_\rho \Delta(B)^2_\rho \geq(\frac{1}{4} \mid<C>\mid^2 + \frac{1}{4}\mid <\big\{A, B\big\}>-2<A> <B>\mid^2)_\rho.
\end{equation} for any density matrix $\rho$, here $<A>_\rho=tr(\rho A)$ is the average of the observable A, and $\Delta(A)^2 = Tr(\rho A^2)-Tr(\rho A)^2$ is its variance (and similarly for observable B).  The partial transpose $\rho^{pt}$ of a bipartite separable density matrix must be positive, which implies that it represents some physical quantum state,  therefore must obey SRUR; \begin{equation}
\Delta(A)^2_{\rho^{pt}} \Delta(B)^2_{\rho^{pt}} \geq (\frac{1}{4} \mid<C>\mid^2  + \frac{1}{4}\mid <\big\{A, B\big\}>-2<A> <B>\mid^2)_{\rho^{pt}}
\end{equation} 
Using, $Tr(\rho^{pt} A)=Tr(\rho A^{pt})$ for any observable A and any density matrix $\rho$, one can shift the partial transposition operation on operators $A, A^2, B, B^2 $ and obtain,
\begin{equation}
 \Delta(A)^2_{pt} \Delta(B)^2_{pt}  \geq \frac{1}{4} \mid<C^{pt}>\mid^2 +  \frac{1}{4}\mid <\big\{A, B\big\}_{pt}>-2<A^{pt}> <B^{pt}>\mid^2. 
\end{equation} 
The above equation still deals with observable quantities as partial transpose of an observable remains an observable.  It is never violated for the separable states and a violation is sufficient to detect entanglement.  We now define suitable observables A and B in order to construct useful uncertainty relation for the entanglement detection,\\
$$A=\frac{1}{2}(\sigma_x\otimes \mathbb{I} + \mathbb{I} \otimes \sigma_x), B=\frac{1}{2}(\sigma_z\otimes\sigma_y + \sigma_y\otimes\sigma_z)$$
with $[A, B] =\imath C$, where $C= \sigma_z\otimes\sigma_z-\sigma_y\otimes\sigma_y$ and $\big\{A, B \big\}=0.$ 
 Partial transposition of A, B and C leads to,\\
 $$A^{pt} = A, B^{pt}= \frac{1}{2}(\sigma_y \otimes \sigma_z - \sigma_z \otimes \sigma_y), C^{pt}=(\sigma_z \otimes \sigma_z + \sigma_y \otimes \sigma_y).$$
 Square of the operators A and B yields,
 $$A^2 = B^2=\frac{1}{2}\begin{bmatrix}1 & 0 & 0 & 1\\ 0& 1 & 1 & 0\\ 0&  1&  1& 0\\ 1&  0&  0& 1
\end{bmatrix}$$ \\ 
 In the operator form, $A^2 = B^2 = \frac{1}{2}(\mathbb{I} + \sigma_{x} \otimes \sigma_{x}).$  It is noted that the $(B^2)^{pt} \neq (B^{pt})^2.$
Under partial transposition,  $(A^2)^{pt}=(B^2)^{pt}=A^2=B^2= D.$  Using  Eq.$[8]$, the separability condition for the density matrix $\rho$ is obtained as,

\begin{equation}
 <D>^2 \geq \frac{1}{4} \mid<C^{pt}>\mid^2 + (<A^{pt}>^2 + <B^{pt}>^2) <D> 
\end{equation} 
If $<A^{pt}>=<B^{pt}>=0,$ Eq.$[9]$ leads to, 
$$ I  + \sigma_{x} \otimes \sigma_{x}  - \sigma_{y} \otimes \sigma_{y}  - \sigma_{z} \otimes \sigma_{z} \geq 0,$$ and $$ I  + \sigma_{x} \otimes \sigma_{x}  + \sigma_{y} \otimes \sigma_{y}  + \sigma_{z} \otimes \sigma_{z} \geq 0,$$which are bi-linear entanglement witness operator $[20]$.  Thus it is clear that  Eq.$[9]$ provides a stronger condition for inseparability than the bi-linear entanglement witness operator.\\
We now explicate the application of the above uncertainty relation.  For the state $\alpha \mid00> - \beta \mid11>,$ using $<A^{pt}>=<B^{pt}>=0$ Eq.$[9]$ yields,\\
$$\frac{1-2\Re(\alpha*\beta)}{2} \geq \frac{1+2\Re(\alpha^*\beta)}{2},$$ which is violated if $\Re(\alpha^*\beta)>0.$  For the Werner states, $\rho_W=x\mid\psi><\psi\mid + \mathbb{I}\frac{(1-x)}{4}$, $$\frac{1-2x\Re(\alpha^*\beta)}{2}\geq \frac{x+2x\Re(\alpha^*\beta)}{2},$$ which is violated if $x > \frac{1}{1+4\Re(\alpha^*\beta)}$.\\  In a particular case, when $\mid\psi>=\frac{\mid00>-\mid11>}{2}$, condition for violation of inequality is  $x>\frac{1}{3},$ the exact condition for entanglement $[9]$.  This result improves the limit of detection given by the Bell inequalities $[21]$ $(x>\frac{1}{\sqrt{2}})$ or, by the uncertainty relations of Guhne $[14]$ $(x>\frac{1}{\sqrt{3}})$ and Gillet $[15]$  $(x>\frac{1}{2})$. These conditions are also valid for the state $\alpha \mid01> - \beta \mid10>,$

In conclusion, we have derived separability inequalities for finite dimensional bipartite systems analogous to the continuous variable quantum system, violation of which provide sufficient condition for entanglement.  They have been obtained by noticing the bounds on the invariants for the finite dimensional bipartite system and from the generalized SRUR, using suitable operators.  Since these inequalities involve Hermitian operators, they can be tested experimentally. We expect that such separability inequalities may be derived for the multipartite quantum systems using appropriate commutation relation and conservation laws.
\section{Acknowledgement}
We acknowledge useful comments from Prof. G. S. Agarwal and Prof. Paul Busch.
{}

\end{document}